\documentclass[aps,prc,twocolumn,showpacs,superscriptaddress]{revtex4-1}
\usepackage{natbib}
\usepackage{graphicx}
\usepackage{color}
\usepackage{amsfonts}
\usepackage{amsmath}
\usepackage{soul}
\usepackage{lineno}
\usepackage{hyphenat}
\newcommand{\heff}{$H_{\rm eff}$}

\newcommand{\qbox}{$\hat{Q}$-box}

\begin{document}
\bibliographystyle{revtex}
\title{Shell-model study of titanium isotopic chain with chiral two- and three-body  forces}
\vspace{0.5cm}

\author{L. Coraggio}
\affiliation{Dipartimento di Matematica e Fisica, Universit$\grave{a}$ degli Studi della Campania "Luigi Vanvitelli",
viale Abramo Lincoln 5, I-81100 Caserta, Italy}
\affiliation{Istituto Nazionale di Fisica Nucleare, Complesso Universitario di Monte S. Angelo, Via Cintia, I-80126 Napoli, Italy}
\author{G. De Gregorio}
\affiliation{Dipartimento di Matematica e Fisica, Universit$\grave{a}$ degli Studi della Campania "Luigi Vanvitelli",
viale Abramo Lincoln 5, I-81100 Caserta, Italy}
\affiliation{Istituto Nazionale di Fisica Nucleare, Complesso Universitario di Monte S. Angelo, Via Cintia, I-80126 Napoli, Italy}
\author{A. Gargano}
\affiliation{Istituto Nazionale di Fisica Nucleare, Complesso Universitario di Monte S. Angelo, Via Cintia, I-80126 Napoli, Italy}
\author{N. Itaco}
\affiliation{Dipartimento di Matematica e Fisica, Universit$\grave{a}$ degli Studi della Campania "Luigi Vanvitelli",
viale Abramo Lincoln 5, I-81100 Caserta, Italy}
\affiliation{Istituto Nazionale di Fisica Nucleare, Complesso Universitario di Monte S. Angelo, Via Cintia, I-80126 Napoli, Italy}
\author{T. Fukui}
\affiliation{RIKEN Nishina Center, Wako 351-0198, Japan}
\author{Y. Z. Ma}
\affiliation{Guangdong Provincial Key Laboratory of Nuclear Science, Institute of Quantum Matter, South China Normal University, Guangzhou 510006, China}
\author{F. R. Xu}
\affiliation{School of Physics, and State Key Laboratory of Nuclear Physics and Technology, Peking University, Beijing 100871, China}

\date{\today}

\begin{abstract}
The even-even Ti isotopic chain,  from $A=42$ to 70,  has been  studied within the nuclear shell-model framework by employing an effective  Hamiltonian which is  derived by way of many-body perturbation theory from a chiral potential with two- and three-body forces, and includes three-body contributions which account for Pauli  principle violations in nuclei with more than two valence particles. We consider $^{40}$Ca as   a closed core and  a model space spanned by  the neutron and proton $0f1p$ orbitals with the addition of the $0g_{9/2}$ orbital for neutrons. 
Calculated two-neutron separation energies and excitation energies of the yrast  $2^+$ states are reported and  compared with the experimental data, which are  available up to $^{62}$Ti.  The present study  intends to investigate  the effects of the adopted  effective interactions on the evolution of the shell structure.

\end{abstract}
\maketitle

\section{Introduction}
\label{intro}
In the last two decades or so, a large amount  of data provided by radioactive ion beam facilities has refined our vision of the nuclear shell structure. The existence of the  magic numbers associated with  2, 8, 20, 28, 50, 82 protons or neutrons, and  126  neutrons has been the cornerstone of the  shell model  for many years starting from its entry into nuclear physics by Mayer and Jensen~\cite{Mayer49,Jensen49}. The new data have put into evidence that  magic numbers are not universal,  but  they may change  as a function of $N$ and $Z$ when going far from the stability line. 
For instance,  new shell gaps have been identified  at $N=32, 34$ in neutron-rich nuclei with $Z\approx 20$ and  at $N=16$ for O isotopes,  while  the disappearance  or weakening of  canonical magic numbers  has been observed in lighter nuclei at $N=8$,  $N=20$, and $N=28$  (see, for instance, Refs.\cite{Huck85,Motobayashi95,Navin00,Iwasaki00,Janssens2002,Shimoura03,Dinca2005,Burger05,Gade06,Bastin07,Hoffman09,Kanungo09,Takeuchi12,Wienholtz13,Steppenbeck13,Liu19}).
Nowadays, great theoretical and experimental efforts are  devoted to study the  evolution of the shell structure and, in particular,  to identify the behavior of shell gaps along isotopic chains as well as to understand  the role played by the different  components of the nuclear force in driving their modifications. 

In  particular,  the Ca isotopic chain, with magic proton number $Z=20$, has attracted considerable interest, since it exhibits two neutron shell closures at $N=32$~\cite{Huck85,Gade06,Wienholtz13} and 34~\cite{Steppenbeck13} in addition to the standard one at $N=28$.   Special attention is presently focused on the very exotic $^{60}$Ca with the aim to acquire information useful to investigate  the shell structure evolution at $N=40$  and shed light on the   possible doubly magic nature of this nucleus.
 From the experimental point of view the existence of $^{60}$Ca has been only recently  established~\cite{Tarasov18},   but no information is available on the excitation energy of the yrast  $2^+ $ state. It is worth mentioning also that  the   experimental location of the neutron drip line in Ca isotopes is still unknown,  and no clear indications emerge from theory \cite{Meng02,Holt12,Hagen12,Hergert14,Neufcourt19,Cao19,Li20}. 

 On the other hand, the shell closure at $N=40$ observed  in
$^{68}$Ni, which corresponds to the filling of the neutron $0f1p$
and proton $0f_{7/2}$ orbitals, rapidly disappears when removing
protons from $0f_{7/2}$, as it is shown by  the behavior of the
experimental excitation energies of the yrast $J^{\pi}=2^+$ states in
iron and chromium neutron-rich isotopes.

In this context, also the chain of Ti isotopes ($Z=22$)  represents a key piece to investigate the  evolution of the shell structure towards $Z=20$. The properties of these nuclei, especially those for  $N\approx40$, may provide a more sound testing ground of nuclear models and of their predictions of the  Ca drip line. 

Experimental data are now known for Ti isotopes from $A= 42$ to 62.  The masses of  $^{60,62}$Ti were determined for the first time  through the measurements reported in Ref. \cite{Michi2020}, while the  $2^{+}_{1}  \rightarrow 0^{+}_{gs}$ and $4^{+}_{1}  \rightarrow 2^{+}_{1} $ transitions in $^{60}$Ti and $^{62}$Ti were  observed  via  one-proton stripping reactions from $^{61}$V and $^{63}$V,  respectively, in Refs. \cite{Gade14,Cortez20}. These experiments show an increase in the   $2^{+}_{1}$ excitation energies of  both nuclei  as compared with the corresponding Cr and Fe isotones, which  indicates a decrease in collectivity towards $Z=20$. However, the reduced size in the increase does not seem to support the existence of  a shell closure at $N=40$ for $^{60}$Ca.

From the theoretical point of view, several shell-model (SM) calculations have been carried out for Ti nuclei  by considering  various  model spaces referred to both $^{40}$Ca or $^{48}$Ca as closed core and employing empirically corrected interactions   \cite{Poves2005,Honma2005,Dinca2005,Lenzi2010,Suzuki2013,Wimmer2019,Gade14,Cortez20}. 
In Ref. \cite{Coraggio2014},  some of the present authors performed a study of isotopic chains  "north-east" of $^{48}$Ca, including Ti isotopes,  by employing two-body matrix elements of the residual interaction derived from the high-precision CD-Bonn free nucleon-nucleon potential.  

 In the present paper,  the  whole even-even titanium isotopic chain from $A=42$ to 70 has been studied  for the first time within the realistic SM framework starting from up-to-date two- and three-body forces derived in the framework of  chiral perturbation theory (ChPT).
 The SM effective Hamiltonian, consisting of the single-particle (SP) energies and the two-body matrix elements (TBMEs), is  built up using  the many-body perturbation theory \cite{Kuo71,Coraggio2020b},    and includes the effect of second-order three-body diagrams arising  in nuclei with more than two valence particles, which account  for three-body correlations induced by the interaction  via the two-body force of clusters of three-valence nucleons
with core excitations as well as with virtual intermediate nucleons scattered above the model space. 
  
This work follows two of our previous studies. In the first one ~\cite{Ma19}, we  investigated the  role played by genuine and effective chiral three-body forces  in providing a reliable monopole component of the SM effective  Hamiltonian defined in the $0f1p$ model space.  Results are discussed for $Z=20$, 22, 24, 26, 28 with $N$ from 22 to 40. In the second one ~\cite{Coraggio20}, focused only on Ca isotopes, the same approach was used in a larger  model space including  the neutron $0g_{9/2}$ orbital  to describe isotopes beyond $N = 40$.  Here, we move a step forward to test our approach along an extensive isotopic chain, as Ti nuclei,  with both valence neutrons and protons.
 
It is  worth mentioning that  two- and three-body  chiral forces have been  also used in some recent {\it ab initio} calculations for Ti isotopes. 
In Ref. \cite{Hagen12}, the authors investigated $^{50,54,56}$Ti by way of the coupled-cluster method  including the coupling to the particle continuum,   while  in Ref. \cite{Xu2019}  the  shell gap in $^{54}$Ti was analyzed in terms of valence-space in-medium similarity renormalization group (VS-IMSRG). The latter  was  also used in Ref.~\cite{Stroberg20} to investigate the  properties of nuclei from helium to iron.
Precision mass measurements of $^{51-55}$Ti were performed in Ref. \cite{Leiste2018}  and compared with the predictions of multireference in-medium similarity renormalization group (MR-IMSRG),  VS-IMSRG, and  self-consistent Gorkov-Green's function (GGF)  calculations.  Furthermore, VS-IMSRG results are reported   in  Ref.~\cite{Cortez20} for the excitation energies of the new observed $2^+$ and $4^+$ states  in   $^{62}$Ti,  together with predictions   of beyond mean-field  and large-scale SM calculations, the latter being based  on the same model space  and effective interaction of Ref.~\cite{Lenzi2010}.  The comparison with experiment shows that only large-scale SM calculations reproduce very accurately data for $^{62}$Ti, which are instead largely overestimated  by the other two approaches. The theoretical framework of Ref. \cite{Lenzi2010} gives also a good description of the experimental excitation energies of $^{60}$Ti, as shown in Ref. \cite{Gade14}.

In concluding this section, it  has to be pointed out that, similarly to  our previous study on Ca isotopes \cite{Coraggio20},  we do not include in the model space the neutron $1d_{5/2}$ orbital, which  was shown to be of   great importance  to reproduce the onset of the collectivity at $N=40$ for isotopic chains "north-east" of $^{48}$Ca, as first discussed  in Ref. \cite{Caurier02} and then confirmed in Refs. \cite{Lenzi2010,Coraggio2014,Wimmer2019,Gade14,Cortez20}.
 
In Ref. \cite{Zuker95}, the authors have identified the source of this
collective behavior as a consequence of the quasi-SU(3) approximate
symmetry, owing to the interplay between the quadrupole-quadrupole
component of the residual interaction and the central field in the
sub-space spanned by the lowest $\Delta j$=2 orbitals of a major
shell.

 In particular, in Ref. \cite{Lenzi2010} -where  the adopted model space was based on a $^{48}$Ca core including the $1p0f$  shell for protons and the $1p,0f_{5/2},0g_{9/2},1d_{5/2}$  orbitals for neutrons - the deformation driving role of the neutron $1d_{5/2}$  orbital below $^{68}$Ni  was assessed showing that the maximum deformation develops in $^{64}$Cr and decreases towards Ca. Such a situation was, in fact, explained in terms of a reduction of the neutron $0f_{5/2} -0g_{9/2}$  gap when protons are removed from the $0f_{7/2}$ orbital, which is accompanied by an enhancement of the quadrupole-quadrupole correlations between the neutron $0g_{9/2}$ and $1d_{5/2}$ orbitals. Similar results are shown in Ref. \cite{Coraggio2014}  where the Ca, Ti, Fe Cr, Ni isotopic chains "north-east"   of $^{48}$Ca were studied  within the  realistic SM framework, by comparing the results  for two model spaces  differing for the inclusion of the  neutron $1d_{5/2}$ orbital.
 
A model space space starting from $^{40}$Ca as closed core and including the neutron $1d_{5/2}$ makes calculations  around $N=40$ very cumbersome.  Dimensions of the Hamiltonian matrix reach about $10^{11}$  in the case of $^{62}$Ti, which represents  the limit of our computing power.
Our choice to exclude  the neutron $1d_{5/2}$  orbital is, therefore, related essentially to this reason. We plan to overcome this limit by extending the so-called double-step procedure introduced in Refs. \cite{Coraggio15,Coraggio16} to reduce the computational complexity of large-scale SM calculations, which consists in  deriving   an effective Hamiltonian within a manageable model space by means of a unitary transformation of a  large-scale Hamiltonian.
 
Nevertheless, the present study, in parallel with our previous one on Ca isotopes, intends to investigate the effects of our SM effective interactions on the evolution of the shell structure  before tackling the problem in a larger model space including the neutron $1d_{5/2}$ orbital.

In the following section,  we give  an outline of the theoretical framework  in which our SM calculations are performed. Results for the excitation energies of the yrast $2^+$ states and two-neutron separation energies of even-even Ti nuclei from $N=20$ to 48 are presented  and compared with the available experimental data in Sec.~\ref{results}. 
In this section,  we also discuss  the  sensitivity of these results with  respect to  the many-body correlations   and  their impact on  the  effective-single particle energies. Section~\ref{conclusions} provides a summary and concluding remarks.

 \section{Outline of the theoretical framework}
\label{theory}

As mentioned in the Introduction, SM calculations for Ti
isotopes have been performed within the same approach as
Ref. \cite{Coraggio20}, which we refer for more details.
Calculations are performed by means of the SM code
KSHELL~\cite{kshell} in the model space spanned by the four proton
$0f_{7/2}$, $0f_{5/2}$, $1p_{3/2}$, $1p_{1/2}$ orbitals and the five
neutron $0f_{7/2}$, $0f_{5/2}$, $1p_{3/2}$, $1p_{1/2}$, $0g_{9/2}$
orbitals  outside the doubly magic  $^{40}$Ca.

We start our calculation from the nucleon-nucleon ($NN$) potential
developed  by Entem and Machleidt \cite{Entem02} within chiral
perturbation theory at next-to-next-to-next-to-leading order (N$^3$LO)
and  the  chiral three-body ($NNN$) potential at
next-to-next-to-leading order (N$^2$LO), which  share the same nonlocal regulator function.
The low-energy constants (LECs) appearing in both  $NN$ and $NNN$
components, namely $c_1$, $c_3$,  $c_4$,  are determined by the
renormalization procedure described in Ref.~\cite{Machleidt11},  while  for
the $c_D$ and $c_E$ LECs characterizing only the $NNN$ force  we  take
the values of Ref.~\cite{Navratil07}.

The matrix elements of the $NN$ and $NNN$ forces, with the addition of
the Coulomb one in the proton-proton channel, are computed in the
harmonic oscillator basis with an oscillator parameter $\hbar
\omega=45A^{-1/3} -25 A^{-2/3}$ for $A=40$.
Details of  the calculation  of matrix elements of the N$^2$LO $NNN$
potential  are reported in Ref.~\cite{Fukui18}.

These matrix elements  are used as input to derive  the  SM effective
Hamiltonian \heff~ within the time-dependent perturbation theory.
Specifically, \heff~ is expressed by way of the Kuo-Lee-Ratcliff
folded-diagram expansion~\cite{Kuo71} in terms of the \qbox~vertex
function, which is defined as

\begin{equation}
\hat{Q} (\epsilon) = P H_1 P + P H_1 Q \frac{1}{\epsilon - Q H Q} Q
H_1 P ~, \label{qbox}
\end{equation}

\noindent
where $H$ is the full nuclear Hamiltonian $H=H_0+H_1$, $H_0$ and $H_1$
being the unperturbed and the interaction components, respectively,
and $\epsilon$  an energy parameter called   starting energy''.

Then, the \qbox~may be calculated by expanding the term $1/(\epsilon -
Q H Q)$ as a power series

\begin{equation}
\frac{1}{\epsilon - Q H Q} = \sum_{n=0}^{\infty} \frac{1}{\epsilon -Q
  H_0 Q} \left( \frac{Q H_1 Q}{\epsilon -Q H_0 Q} \right)^{n} ~.
\end{equation}

\noindent
This provides a perturbative expansion of the \qbox, and its
diagrammatic representation is given as a collection of irreducible
valence-linked Goldstone diagrams \cite{Kuo71}.

Once the \qbox~is calculated, \heff~is obtained solving nonlinear
matrix equations by way of iterative techniques such as the
Kuo-Krenciglowa and Lee-Suzuki ones \cite{Suzuki80}, or
graphical noniterative methods \cite{Suzuki11}.
The latter is the method we have employed in present work, since it
results in a faster and more stable convergence to the solution of the
matrix equations.

We arrest the $\hat{Q}$-box expansion of the one- and two-body
Goldstone diagrams at third order in the $NN$ potential and at first
order in the $NNN$ one.
It has to be pointed out that the diagrams at first order in $NNN$
potential - whose analytical expressions are reported in
Refs. \cite{Fukui18,Ma19} - are the coefficients of the one-body and
two-body terms arising from the normal-ordering decomposition of the
three-body component of a many-body Hamiltonian \cite{HjorthJensen17}.

 Since our goal is the study of the Ti isotopic chain up
  to $^{70}$Ti, we are going to diagonalize the SM Hamiltonian for
  systems up to thirty valence nucleons.
  This means that the derivation of the SM effective Hamiltonian needs
  to resolve the progressive  filling of the model space orbitals,
  especially in the calculation of the irreducible valence-linked
  diagrams of the \qbox.

\begin{figure}[h]
\begin{center}
\includegraphics[width=6cm]{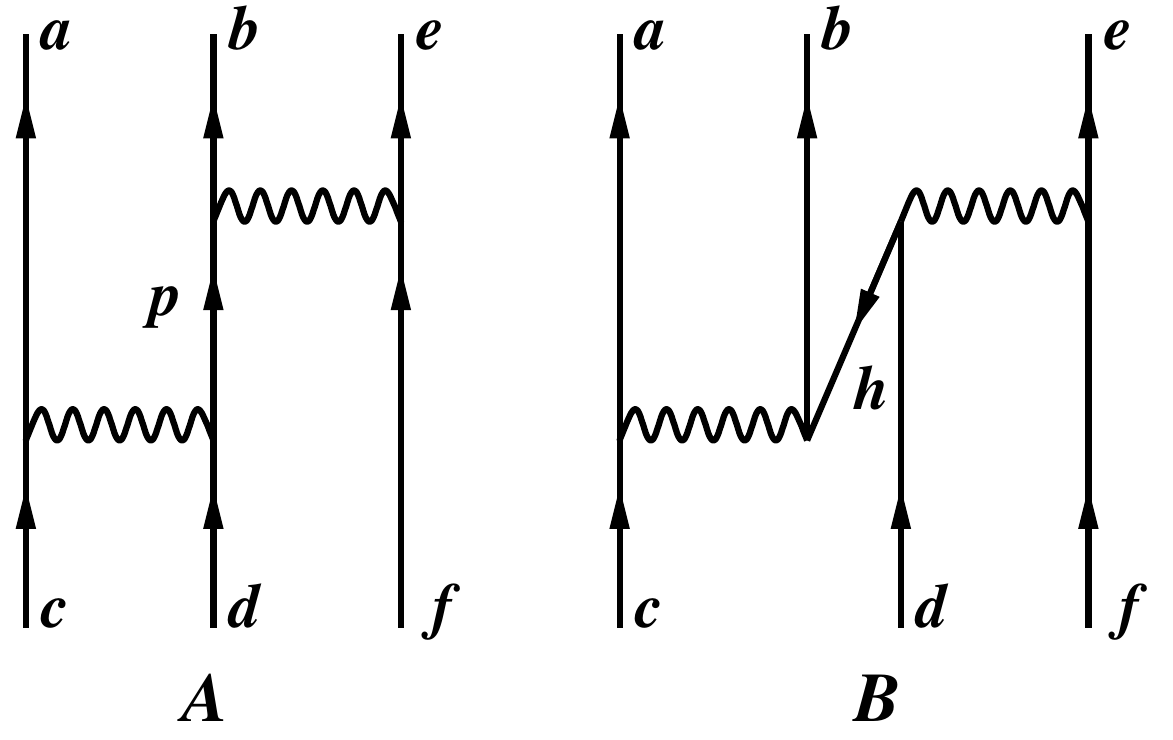}
\caption{Second-order three-body diagrams. The sum over the
  intermediate lines runs over particle and hole states outside the
  model space, shown by $A$ and $B$, respectively. We report only one
  of the nine existing topologies which correspond to the permutations of
  the external lines.}
\label{diagram3corr}
\end{center}
\end{figure}

  This should be achieved by including in the \qbox~many-body diagrams
  which account for the interaction via the two-body force of the
  valence nucleons with configurations outside the model space,
  leading to a dependence of \heff~on the number of valence nucleons.
  We arrest the cluster expansion to the leading term, namely
  second-order three-body diagrams, which, for those nuclei with more
  than two valence nucleons, account for the interaction of the valence
  nucleons with core excitations as well as with virtual intermediate
  nucleons scattered above the model space.
  These diagrams are reported in Fig. \ref{diagram3corr} and their
  explicit expressions, $D_A$ and $D_B$,  are 
  
\begin{multline}\label{DA}
\langle \left[ (j_{a} j_{b})_{J_{ab}},j_e \right]_{J} | D_A  | \left[
  (j_{c} j_{d})_{J_{cd}},j_f \right]_{J} \rangle  = \\
\sum_{J_{be}} \sum_{p} (-1)^{b+e+f+p} \hat{J_{ab}} \hat{J_{cd}} \hat{J_{be}}^2
\left\{ \begin{array}{ccc}
j_a~ j_b~  J_{ab}  \\
 j_e~  J~  J_{be}  \end{array} \right\}
\left\{ \begin{array}{ccc}
j_a~ j_p~  J_{cd}  \\ 
j_f~  J~  J_{be}  \end{array} \right\} 
\\
~ \times \frac{
\langle b,e; J_{be}| V_{NN} | p,f; J_{be}\rangle
\langle a,p; J_{cd}| V_{NN} | c,d; J_{cd}\rangle
}
{
[\epsilon_{0}-(\epsilon_{a}+\epsilon_{f}+\epsilon_{p})]
}~, 
\end{multline} 

\begin{multline}\label{DB}
\langle \left[ (j_{a} j_{b})_{J_{ab}},j_e \right]_{J} | D_B  | \left[
  (j_{c} j_{d})_{J_{cd}},j_f \right]_{J} \rangle  = \\
\sum_{J_{be}} \sum_{h} (-1)^{b+e+f+h+1} \hat{J_{ab}} \hat{J_{cd}} \hat{J_{be}}^2
\left\{ \begin{array}{ccc}
j_a~ j_b~  J_{ab}  \\
 j_e~  J~  J_{be}  \end{array} \right\}
\left\{ \begin{array}{ccc}
j_a~ j_h~  J_{cd}  \\ 
j_f~  J~  J_{be}  \end{array} \right\} 
\\
~ \times \frac{
\langle b,e; J_{be}| V_{NN} | h,f; J_{be}\rangle
\langle a,h; J_{cd}| V_{NN} | c,d; J_{cd}\rangle
}
{
[\epsilon_{0}-(\epsilon_{a}+\epsilon_{b}+\epsilon_{d}+\epsilon_{f}-\epsilon_{h})]}~,
\end{multline} 

\noindent
where the indices $a,b,c,d,e,f,p,h$ refer to the quantum numbers of
the incoming, outcoming, and intermediate single-particle states,
$\epsilon_m$ denotes the unperturbed single-particle energy of the
orbital $j_m$, $\epsilon_{0}$ is the so-called starting energy, namely
the unperturbed energy of the incoming particles
$\epsilon_0=\epsilon_{c}+\epsilon_{d}+\epsilon_{f}$.

From now on, we dub these contributions   3-b correlations  .

We also point out that for each topology reported in
Fig. \ref{diagram3corr}, there are nine diagrams, corresponding to the
possible permutations of the external lines \cite{Polls83}.

Diagrams $A,B$ include the effects of the
Pauli blocking due to the filling of the valence particle lines in the
second-order ladder and core-polarization diagrams \cite{Ellis77},
respectively, quenching the contribution of these two-body terms.

\begin{figure}[h]
\begin{center}
\includegraphics[width=5.5cm]{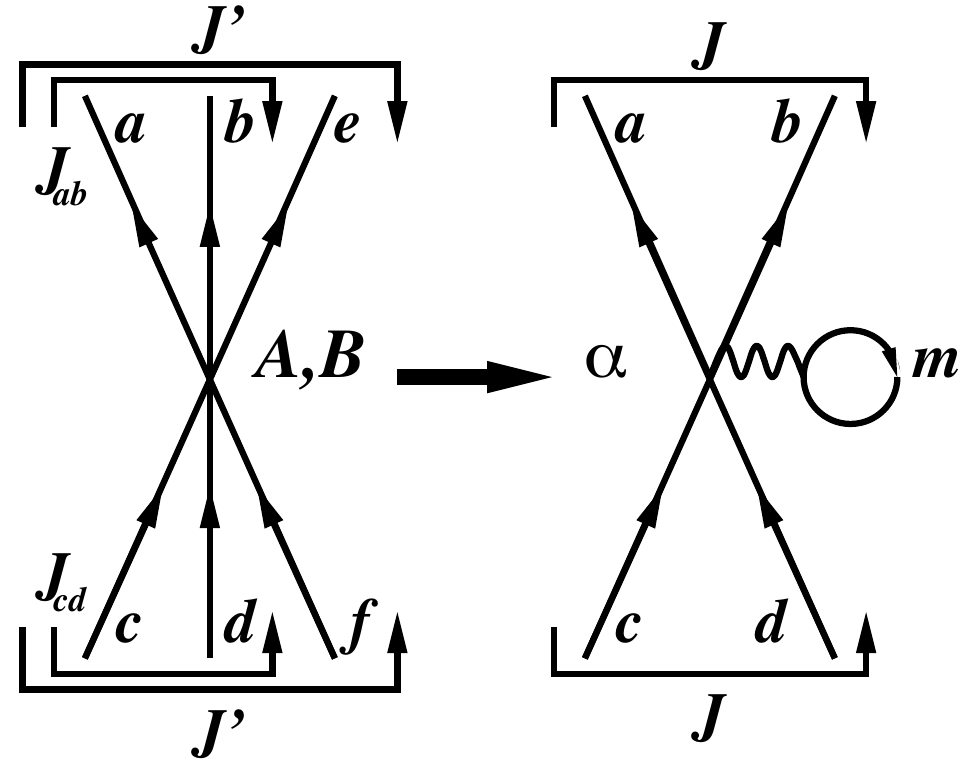}
\caption{Density-dependent two-body contribution that is derived from
  a three-body one. $\alpha$ is obtained by summing over one incoming
  and outgoing particle of the three-body graphs $A,B$ reported in
  Fig. \ref{diagram3corr}.}
\label{2ndord}
\end{center}
\end{figure}

Nevertheless, the KSHELL SM code, which we employ for the
calculations, cannot perform the diagonalization of a three-body
\heff, so we derive  density-dependent two-body terms ($\alpha$) from
the corresponding second-order three-body diagrams  ($A,B$),  as shown
in Fig. \ref{2ndord}.

We then calculate, for each $A,B$ topology, nine one-loop
diagrams, namely the graph $(\alpha)$ in Fig. \ref{2ndord}.
Their explicit form, in terms of the three-body graphs $A,B$, is

\begin{multline}
\label{correq}
\langle (j_a j_b)_J| V^{\alpha} | (j_c j_d)_J \rangle = \\
\sum_{m,J'} ~ \rho_m \frac{\hat{J'}^2}{\hat{J}^2} \langle \left[
  (j_{a} j_{b})_{J},j_m \right]_{J'} | V^{A,B}  | \left[
  (j_{c} j_{d})_{J},j_m \right]_{J'} \rangle~~,
\end{multline}

\noindent
where the summation over $m$-index runs in the model space, and
$\rho_m$ is the unperturbed occupation density of the orbital $m$
according to the number of valence nucleons.

Following the above procedure, the perturbative expansion of the
$\hat{Q}$-box contains one- and two-body diagrams up to third order in
$V_{NN}$, and a density-dependent two-body contribution which accounts
for three-body second-order diagrams $A,B$.

It should be stressed that the latter term depends on the number of
valence protons and neutrons, thus we derive specific effective
SM Hamiltonians for any nuclear system under consideration,
Hamiltonians that differ only for the two-body matrix elements.

It is worth mentioning that, as done in Ref. \cite{Coraggio20} for
Ca, we have checked the effect of the spurious center-of-mass motion.
In line with the previous outcome, we have found that results are only
marginally affected by  spurious components.

In the Supplemental  Material~\cite{SM} we report the TBMEs  of the
interaction calculated without including three-body correlations as
well as those of the density dependent interaction for $A=62$.
Neutron and proton SP energies are  shown in Table~\ref{tab-1}.
It is worth mentioning that the proton and neutron SP spacings
obtained from the theory are shifted to reproduce the experimental
ground-state energies of $^{41}$Sc and $^{41}$Ca with respect to
$^{40}$Ca.
This is because, as discussed in Ref. \cite{Coraggio20}, our derivation of
the effective Hamiltonian with $N_{\rm max}=18$ does not provide
convergent spectra for the one-valence systems.

\begin{table}[h]
\begin{ruledtabular}
\caption{Proton $\epsilon_{b}^{\pi}$ and neutron $\epsilon^{\nu}_{b}$
  single-particle energies (in MeV).}
\label{tab-1}
\begin{tabular}{c r  r } 
$b $& $\epsilon^{\pi}_{b}$ & $\epsilon^{\nu}_{b}$  \\
\hline
$0f_{7/2} $& $-1.1$ & $-8.4$ \\ 
$0f_{5/2} $& $6.2$ & $-0.2$ \\ 
$1p_{3/2}$ & $1.4$ & $-5.2$ \\ 
$1p_{1/2}$ & $3.3$ & $-3.1$ \\
$0g_{9/2}$ & $-$ & $1.5$ \\ 
\end{tabular}
\end{ruledtabular}
\end{table}

\section{Results}
\label{results}
\begin{figure}[ht]
\includegraphics[width=\columnwidth ]{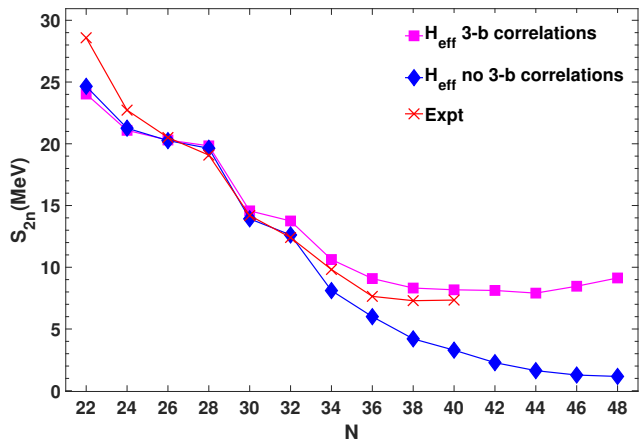}  
\caption{(Color online) \label{fig1} Theoretical two-neutron separation energies for even-even Ti isotopes from $N=22$ to 48,  with and without the  3-b correlations, are  compared to  the experimental data from Ref. \cite{Michi2020} for $N=40$ and  from AME2020 \cite{AME2020} for the other isotopes.} 
\end{figure}

We start  focusing  on the two-neutron separation energies, whose calculated values from $^{44}$Ti to $^{70}$Ti are  shown in Fig.~{\ref{fig1}} together with the available experimental data.  
 
We see that the   calculated two-neutron separation energies ($S_{2n}$),  with (magenta squares) and without (blue diamonds) 3-b correlations, are in a good agreement with the experimental data (red cross)  up to $N=32$. Actually, both calculations provide the observed shell closures at  $N=28$ and $N=32$, which manifest as a sudden drop of $S_{2n}$ at $N=30$ and $N=34$. 
However, starting from $N=34$ the effect of  3-b correlations becomes rather large and, as expected, grows with increasing neutron number  in analogy with the results obtained  for Ca isotopes \cite{Coraggio20}.  
It turns out that their contribution  increases the calculated  $S_{2n}$,  and leads  to a good agreement with the  measured values in $^{56-62}$Ti,  which are  significantly  underestimated when 3-b correlations are omitted.
Nevertheless,  in both calculations  all  Ti isotopes up to $^{70}$Ti are predicted to be bound.

\begin{figure}[ht]
\includegraphics[width=\columnwidth ]{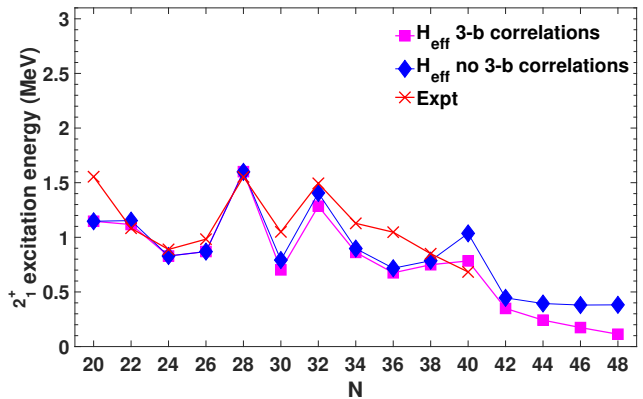}  
\caption{(Color online) \label{fig2} Theoretical $2^+_1$ excitation energies for even-even Ti isotopes from $N=20$ to 48, with and without the  3-b correlations, are compared with the experimental data from Ref. \cite{Cortez20} for $N=40$ and  from Ref. \cite{NNDC} for the other isotopes.}
\end{figure}

The experimental excitation energies of the yrast $2^+$ state are compared  with the calculated values obtained with and without  3-b correlations in Fig.~\ref{fig2}. Up to $N=38$, the results of the two calculations are very similar  and do not differ significantly from those reported in Ref.~\cite{Ma19}, where the model space was limited to the neutron and proton $1f1p$ orbitals. It can be seen that  both calculations are in quite good agreement with experimental data up to $N=34$,  and in particular predict the observed shell closures in $^{50}$Ti and $^{52}$Ti.  
 However, from $N=36$ on, the theoretical curves start to climb towards $N=40$, at variance with the observed behavior of the yrast $2^+$ excitation energies that does not exhibit such a raise. 
This deficiency in our calculations may be ascribed to the  inadequacy of the adopted model space, which does not include the neutron $1d_{5/2}$ orbital.
In fact, as mentioned in the Introduction, the interaction between neutron $1d_{5/2}$ and  $0g_{9/2}$ orbitals ignites a quadrupole collectivity that is responsible for the disappearance of the $N=40$ shell closure in isotopic chains "north-east"   of $^{48}$Ca, especially in chromium and iron isotopes \cite{Caurier02,Lenzi2010,Coraggio2014,Wimmer2019,Gade14,Cortez20}.

 Despite this unsatisfactory result, it is worth pointing out that differences emerge in the two theoretical curves starting from $N=40$  to 48.
As a matter of fact, they evince that the contribution of 3-b correlations reduces the excitation energy of the $2^{+}_{1}$ state, leading to the disappearance of the shell gap in $^{62}$Ti, in agreement with experiment  \cite{Cortez20}.

We have found that 3-b correlations  also influence the calculation of the yrast $4^+$ states. The predicted  behavior of their excitation energies, obtained by omitting these correlations, exhibits a positive slope from $^{60}$Ti to $^{62}$Ti that is not experimentally observed.
On the other hand, the experimental behavior is reproduced when 3-b correlations are included, although the calculated excitation energies underestimate the measured values by about 500 KeV along the whole chain, except for $^{42}$Ti where a larger discrepancy is noticed.

The results shown above evince the effects produced on the shell structure of Ti isotopes when the evolution the SM Hamiltonian, as a function of the number of valence nucleons, is microscopically taken into account.  We are confident that the combination of these correlations with the enlargement of the model space including the neutron $1d_{5/2}$ orbital is the key to reproduce the observed degree of collectivity in N$\approx$40 nuclei "north-east"   of $^{48}$Ca.

The shell evolution can be analyzed by studying the behavior of the effective single particle energies (ESPEs) along the isotopic chain, and thus we have found it interesting to see how they are influenced by 3-b correlations.  

The ESPEs  are  defined as 
\begin{equation}
{\rm ESPE} (a^\tau)= \epsilon_{a}^{\tau} +  \sum_{b \tau^{\prime}}   {\bar V}_ {a b}^{\tau \tau^{\prime}} n_{b}^{\tau^{\prime}},
\end{equation}
where $\tau$,$\tau^{\prime}$ stand for neutron or proton index and $a$,$b$ run over all the valence orbitals.  The quantities $\epsilon_{a}^{\tau}$ and  $n_{a}^{\tau}$ denote  the   SP energy and the  ground-state occupation number of the $a^\tau$ level, while  ${\bar V}_{a b}^{\tau  \tau^{\prime}}$ is the monopole component of the two-body effective interaction
 $(V_{\rm eff})$ obtained by averaging on the projection of the total angular momentum and can be written as
 
\begin{equation}
{\bar V}_{a b}^{\tau \tau^{\prime}}=	\frac{ \sum_J(2J+1) \langle a^{\tau}, b^{\tau^{\prime}} \mid V_{\rm eff} \mid  a^{\tau}, b^{\tau^{\prime}}  \rangle_J}{\sum_J (2J+1)},
\end{equation}

{\noindent where  $J$ runs over Pauli allowed values.}

The proton  ESPEs  are not significantly affected by   3-b correlations. In both calculations,  the proton $0f^\pi_{7/2}$ orbital is well separated from the other ones, and the  $1p^\pi_{3/2}- 0f^\pi_{7 /2}$ gap increases towards neutron rich Ti isotopes, which indicates the stability of the shell closure at 
$Z  = 28$. The  other SP states are  close each other with the inversion of $1p^\pi_{1/2}$ and $0f^\pi_{5/2}$ orbitals in correspondence of $N=40$.

\begin{figure}[ht]
\includegraphics[width=\columnwidth ]{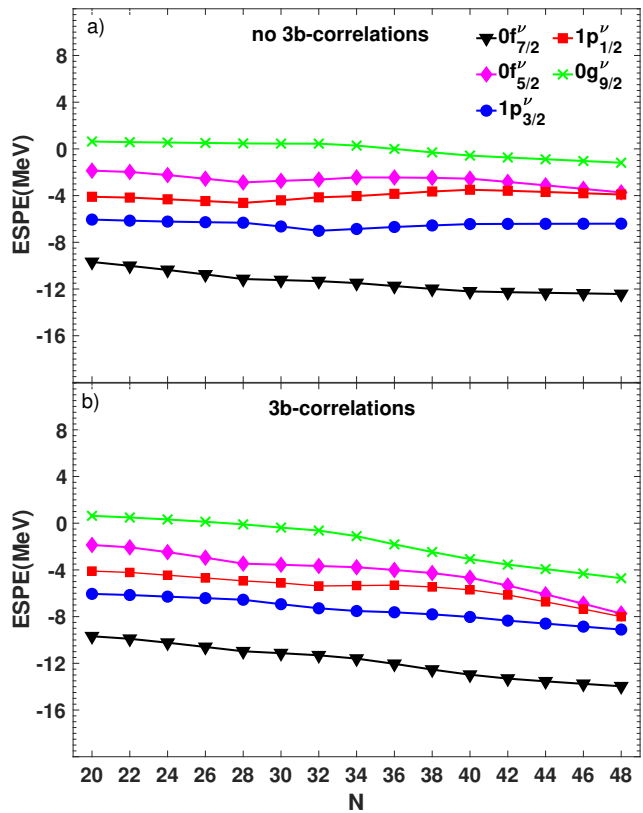}  
\caption{(Color online) \label{fig4} ESPE for neutrons in even-even Ti isotopes  from $N=20$ to 48 resulting from calculations (a) without and (b) with  3-b correlations.}   
\end{figure}

More interesting is the behavior of the ESPEs for  neutrons.
Figure \ref{fig4}(a), presenting the neutron ESPEs obtained by omitting  3-b correlations, shows some similarities with Fig. \ref{fig4}(b), where these correlations are included.  In both calculations, the neutron   $0f_{7/2}$ orbital is well isolated from the others along all the isotopic chain,  and  a fairly large spacing is observed between the  $1p_{3/2}$  and $1p_{1/2}$ orbitals up to $N  = 44$, which gives rise to the subshell closure at $N  = 28$ and $N  = 32$, respectively. No shell closure is instead predicted at $N  = 34$, corresponding to the filling of the $1p^\nu_ {1/2}$ orbital, whose separation energy from the  $0f^\nu_{5/2}$ orbital decreases with increasing neutron number. 
In spite of these similarities, however, we note that 3-b correlations introduce appreciable changes in the neutron ESPEs, which explain the differences in the two-neutron separation energies and in the $2^{+}_1$ excitation energies discussed above.  In particular, we see that all the neutron ESPEs show a more rapid and almost continuous  decrease as a function of $N$ when  3-b correlations are taken into account, while  the neglect of these correlations leads to a more flat behavior or even to an increasing trend  for the $1p^\nu_{3/2}$,  $1p^\nu_{1/2}$, and  $0f^\nu_{5/2}$ orbitals from $N=30$ or $34$ to 40. This explains the strong negative slope of the  $S_{2n}$ curve when calculations are performed without including  3-b correlations.
Moreover, the $0g^\nu_{9/2} - 0f^\nu_{5/2}$ gap   is affected by  3-b correlations which reduce its value by $\approx 400$ keV at $N=40$. The larger gap we find when  3-b correlations are neglected leads to the  increase in energy of  the  $2^+_1$ state from  $^{60}$Ti to $^{62}$Ti we predict in such  a case.

\begin{figure}[ht]
\includegraphics[width=\columnwidth ]{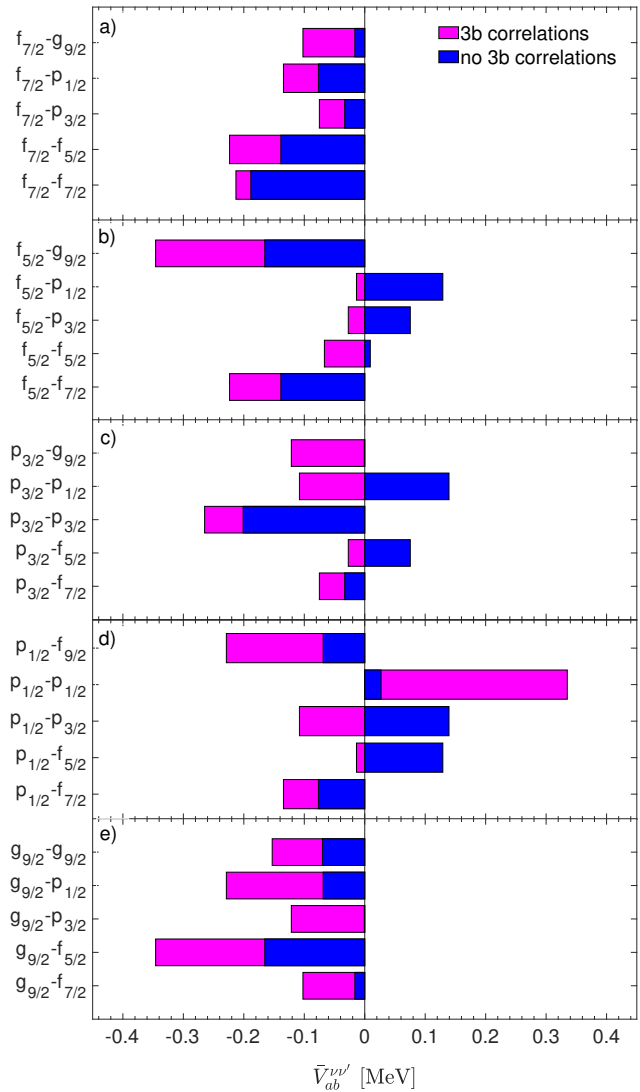}  
\caption{(Color online) \label{fig5}  Neutron-neutron monopole matrix elements, ${\bar V}^{\nu \nu'} _{ab}$, of the effective interactions with and without 3-b correlations (see text for details). }   
\end{figure}

The different behavior of the neutron ESPEs with and without  3-b correlations is related to the changes these correlations produce in the
monopole part of the neutron-neutron interaction. Actually, the neutron-proton monopole components, which also enter in the  neutron ESPE
definition, are not significantly affected  by  3-b correlations, and, additionally, they do  not play a relevant  role since their contribution is limited  by
the reduced number of valence  protons.  As concerns the  neutron-neutron monopole components, we find  that  3-b correlations induce on the overall  a larger attractiveness which
explains the difference in the behavior of the neutron ESPEs shown in Figs. \ref{fig4}(a) and \ref{fig4}(b).  As an example,  we compare in Fig. \ref{fig5} the  monopole matrix elements of  the neutron-neutron effective interaction without 3-b correlations with those of the density-dependent interaction at $N=40$. It can be seen that the only matrix element  that acquire a repulsive component is   ${\bar V}_{1p_{1/2} 1p_{1/2}}^{\nu \nu}$.

Furthermore, the size of changes induced by  3-b correlations is orbital dependent, which leads  to  modifications of the ESPE spacings. In particular, ${\bar V}_{0g_{9/2} b}^{\nu \nu}$  receive a larger attractive contribution   with respect to ${\bar V}_{0f_{5/2} b}^{\nu \nu}$ 
 for any value of $b$, except  for $b=0g_{9/2}$.  This gives rise to the reduced
gap   between the $0_{9/2}$ and $0f_{5/2}$ ESPEs  we find at  $N=40$  when  3-b correlations are taken into account.

To better asses the role of the monopole contributions arising from 3-b correlations we have investigated the interplay of monopole and multipole components. To this end, we have built a modified effective interaction  for each Ti isotope starting from the  Hamiltonian without 3-b correlations and replacing its monopole component with that arising from the  density-dependent Hamiltonian derived for  the same nucleus. The obtained results are very close to those of the original density-dependent Hamiltonians, showing, in particular, that the behavior of the two-neutron separation energies and of the $2^{+}_1$ excitation energies is essentially determined by the monopole component.

 \section{Summary and conclusions}
 \label{conclusions}
In this paper we have  studied the even-even Ti isotopic chain  from $A=42$ to 70   within the realistic SM framework starting from up-to-date two- and three-body forces derived from ChPT.  The  one and two-body matrix components of the SM effective  Hamiltonian are derived by means of the many-body perturbation in a model space spanned by the proton and neutron  $0f1p$ orbitals plus the neutron $0g_{9/2}$ orbital. Three-body contributions are  also included through density-dependent TBME consistently derived within a microscopic approach from chiral forces.  These correlations are essential to account for the Pauli blocking effect due to the progressive filling  of the model space orbitals in nuclei with more than two valence nucleons. 

As in our previous paper~\cite{Coraggio20},  we have compared the calculated  two-neutron separation  energies and the excitation energies of the yrast $2^+$ state  with the available experimental data.

We have found that  3-b correlations have a tiny impact on the two-neutron separation energies up to $N=32$ but, starting from $N = 34$ on, their role becomes more relevant with  the increasing  number of  valence neutrons.
This  results in an upshift of  the two-neutron separation energies that become very close to the experimental ones up   $^{62}$Ti beyond which no experimental information is available. This makes us confident in our predictions for the heaviest Ti isotopes, which we found to be bound up to $N=70$. In Ref. \cite{Coraggio20}, we have shown that 3-b correlations are  relevant  also to
determine the drip line in Ca isotopes. In fact, without the inclusions of these 
 correlations the drip line of calcium isotopes is located at $N = 40$, while their attractive contribution shifts the last bound
nucleus at least to $^{68}$Ca.

As concerns the excitation energies of the yrast  $2^+$ states, we do not obtain a satisfactory agreement with experiment from $N=36$ on, despite the inclusion of 3-b correlations.
 As a matter of fact, the  decreasing trend  observed from $N=36$ to $40$ is not reproduced even if correlations are taken into account.  This may be related to the omission  of the neutron $1d_{5/2}$ orbital, which, as discussed above, is fundamental to reproduce the collectivity   of nuclei "north-east"   of $^{48}$Ca around a $N = 40$, especially of $^{64}$Cr  and $^{62}$Fe. Nevertheless, our calculations evince that contributions arising from 3-b correlations  affect   the neutron-neutron monopole components leading to a reduction  of  the excitation energy of the $2^{+}_{1}$ state and in particular of the shell gap at $^{62}$Ti. A similar result was found for the excitation energy of $2^+_1$  state in $^{60}$Ca, whose value is reduced by  their  inclusion from $\sim 2.3$ to 1.7 MeV, which is quite small when compared with the  gaps    observed in $^{52}$Ca and $^{48}$Ca, namely $\approx 2.5$  and  $\approx 4.0$ MeV. 

Despite the missing contribution coming from the  neutron $1d_{5/2}$  orbital, interesting conclusions can be drawn. In fact, the present study shows that  starting from $N=34$ the  attractive contribution arising from 3-b correlations impacts significantly on the shell  structure evolution.

We plan in a near future to overcome the computational complexity related to the large model by including the neutron $1d_{5/2}$ orbital so as to perform large-scale SM calculations to study the collective behavior at $N=40$ starting from chiral two- and three-body forces and employing single-particle energies and two-body matrix elements derived from many-body perturbation theory.

\begin{acknowledgments}
We acknowledge
the CINECA award under the ISCRA initiative code HP10B51E4M and  through the INFN-CINECA agreement for the availability
of high performance computing resources and support. G. De Gregorio acknowledges the support by the funding program
  VALERE   of Universit\`{a} degli Studi della Campania   Luigi
Vanvitelli  . The National Natural Science Foundation of China under Grants No. 11835001, 11921006, and 12035001 is also acknowledged.
\end{acknowledgments}

\bibliographystyle{apsrev}
\bibliography{paper_Ti}
\end{document}